\documentclass[%
 aps,
 prb,%
 amsmath,amssymb,
preprint,%
superscriptaddress,showpacs]{revtex4-1}

\usepackage{color, graphicx}
\usepackage{dcolumn}
\usepackage{bm}
\usepackage{amssymb}
\usepackage{latexsym}
\usepackage{amsfonts}
\usepackage{amsmath}
\usepackage{multirow}
\usepackage{ifthen}

\begin{document}

\title{Analysis of the wave propagation properties of a periodic array of rigid cylinders perpendicular to a finite impedance surface}

\author{V. Romero-Garc\'ia}
 \affiliation{Instituto de Investigaci\'on para la Gesti\'on Integrada de zonas Costeras, Universidad Polit\'ecnica de Valencia, Paraninf 1, 46730, Gandia, Spain}
 \email{virogar1@gmail.com}
 
 \author{J.V. S\'anchez-P\'erez}
 \affiliation{Centro de Tecnolog\'ias F\'isicas: Ac\'ustica, Materiales y Astrof\'isica, Universidad Polit\'ecnica de Valencia, Camino de Vera s/n, 46022, Valencia, Spain}
 
 \author{L.M. Garcia-Raffi}
 \affiliation{Instituto Universitario de Matem\'atica Pura y Aplicada, Universidad Polit\'ecnica de Valencia, Camino de Vera s/n, 46022, Valencia, Spain}

\begin{abstract}
The effect of the presence of a finite impedance surface on the wave
propagation properties of a two-dimensional periodic array of rigid
cylinders with their axes perpendicular to the surface is both
numerically and experimentally analyzed in this work. In this
realistic situation both the incident and the scattered waves
interact with these two elements, the surface and the array. The
interaction between the excess attenuation effect, due to the
destructive interference produced by the superposition of the
incident wave and the reflected one by the surface, and the bandgap, due to the periodicity of the array, is fundamental for the design
of devices to control the transmission of waves based on periodic
arrays. The most
obvious application is perhaps the design of Sonic Crystals Noise
Barriers. 
  Two different finite impedance surfaces have been analyzed
in the work in order to observe the dependence of the wave
propagation properties on the impedance of the surface.
\end{abstract}

\pacs{43.20.+g, 43.35.+d}

\maketitle

\section{Introduction}
Research on the transmission properties of Phononic Crystals (PC),\cite{Kushwaha93, Sigalas93}
defined as periodic lattices of elastic scatterers embedded in an elastic medium with
different physical properties, have attracted increasing interest in
last decade. Exhaustive studies have been performed in order to know
the underlying physics in their main control properties of sonic and elastic
waves. Refraction,\cite{Cervera02,Yang04,Feng05} invisibility,\cite{Cummer07} bandgaps,\cite{Sanchez98, Sanchez02} or recently,
control of phase properties,\cite{Swinteck11} are some topics related with the exploitation of the 
wave propagation properties of PC. In these years, several theoretical methods
have been developed to analyze the physical properties of these
systems. Plane Wave Expansion method (PWE) \cite{Kushwaha94} allow to find the
bands structure of the PC transforming the wave equation into an
eigenvalue problem that can be solved for each Bloch vector $k$, in
the irreducible first Brillouin zone. Extended Plane Wave Expansion
(EPWE) \cite{Romero10b, Laude09} is a general method for analyzing the complex
dispersion relation of PC. Multiple Scattering Theory (MST) \cite{martin06, linton01, Chen01} is a self
consistent method to estimate the acoustic pressure in a point
taking into account the incident pressure on the PC and the
scattering due to all the scatterers. Finally, Finite Elements Method (FEM) \cite{ihlenburg98} can be used to
characterize periodic media where the geometries of the scatterers or
the existence of some new effects different than scattering are
introduced in the problem. 

In a real situation, scatterers should be hold on a finite impedance surface and the reflections of both, the incident and the scattered waves on this surface have to be taken into account when the transmision properties of PC are analyzed. However, this effect has not been considered in the previous models. As a first approximation to this problem, recent works \cite{Romero11, Krynkin11 } present both semi analytical and numerical models to
analyze the influence of a finite impedance surface on the transmission properties of
arrays of rigid cylindrical scatterers embedded in air, usually called sonic crystals (SC), with their axes
parallels to the surface (horizontal scatterers).  This situation has the most tractable geometry to develop a semi analytic model. However, as we have mentioned, the real case involves arrays of cylinders with their axes
perpendicular to the surface which means a three dimensional (3D)
geometry problem. This case implies to assume the complete geometry
of the composite resulting in very complex analytical developments. On the other
hand, numerical simulations of these 3D geometries would imply a
long computational time. 

The goal of this work is to present
numerical predictions and experimental results to show the variation of the transmission
properties of SC devices when the cylinders are perpendicular to the finite impedance surface. To do this we develop a simplified 3D numerical model based on FEM and we check the results with accurate experimental results  obtained in controlled conditions. The
interaction between the excess attenuation effect, due to the
destructive interference produced by the superposition of the
incident wave and the wave reflected by the surface, and the bandgap, due to the periodicity of the array, has been found of fundamental interest for the design of devices based on periodic arrays devoted to control the transmission of waves. The study is focused
on the airborne transmission case. The results obtained here allow
to enhance the knowledge about the sonic behaviour of these systems
in a more realistic way.

The work is  organized as follows: First of all we explain the used experimental set up and the developed numerical model.
Then a discussion about the obtained numerical results and a
comparison with the experimental ones is shown. Finally, the concluding remarks show the main conclusions of the work.

\section{Experimental set up and Numerical analysis}
\label{sec:tools}

\subsection{Experimental set up}
\label{sec:experimental}
\begin{figure}
\begin{center}
\includegraphics[width=8.5cm]{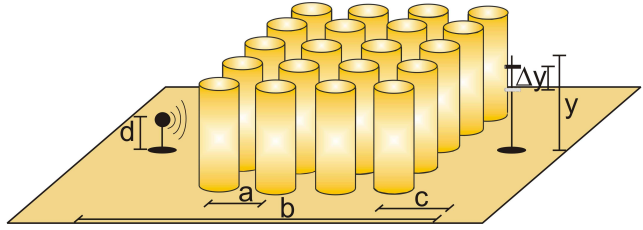}
\end{center}
 \caption{\label{fig:figure2} (Color online) Experimental set up
used}.
\end{figure}

The experimental results have been obtained under controlled
conditions in an echo-free chamber. In fig. \ref{fig:figure2} one can observe the
experimental set up used. We have constructed a SC made of 20
(5x4) acoustically rigid cylinders with radius $r=0.09$ m and $1.20$ m
length, arranged in a square array with lattice constant $a=0.22$ m.
With this geometry  the first pseudogap at $\Gamma$X direction
appears between 500 and 980 Hz, as we will explain later.

 Two types of surfaces have been considered. The first one is a $0.03$ m
thick wooden board, theoretically considered as an acoustic rigid surface. The
second one is a $0.1$ m foam sheet, considered as a porous material with
finite impedance characterized using a two-parameter
impedance model \cite{Taherzadeh99} with flow resistivity $\sigma$=250 kPa s/m$^2$
and porosity at the surface $\alpha=1$ m$^{-1}$. To experimentally study the interaction between the SC
and the surface, we have taken measurements on a vertical line of
points with length $y=1$ m (starting at 0.05 m from the surface)
behind to the SC, as one can see in fig. \ref{fig:figure2}. We have measured this
line in steps of $\Delta y=0.02$ m in order to analyze in-depth
the interaction between the surfaces and the bandgaps of SC by
varying the height of the microphone. The distance between the
vertical line of measurements and the center of the nearest row of
cylinders is $c=0.5$ m, and the horizontal distance between the sound source and the measurement line is $b=2.51$ m. A
source emitting white noise is placed at $d=0.3$ m distance from the
ground.

In our experimental set up, the acquisition is done by a
prepolarized free-field microphone 1/2'' Type 4189 B\&K. The
microphone position is varied by using a Cartesian robot which controls the movement along the three axes (OX, OY and OZ), installed in the ceiling of the echo-free chamber. The robot is called
by us 3DReAMS (3D Robotized e-Acoustic Measurement System), and it has
been designed to sweep the microphone through a 3D grid of measuring
points located at any trajectory inside the chamber. When both the
acoustic source and the robotized system are turned off, the
microphone acquires the temporal signal. This signal is saved on the
computer and then, using the Fast Fourier Transform (FFT), one can
obtain the power spectra, the frequency response or the sound level
measurement. National Instruments cards PCI-4474 and PCI-7334
have been used together with the Sound and Vibration Toolkit and the
Order Analysis Toolkit for LabVIEW for both the data acquisition and
the motion of the robot.

\subsection{Numerical analysis: Finite Element Method}

\begin{figure}
\begin{center}
\includegraphics[width=8.5cm]{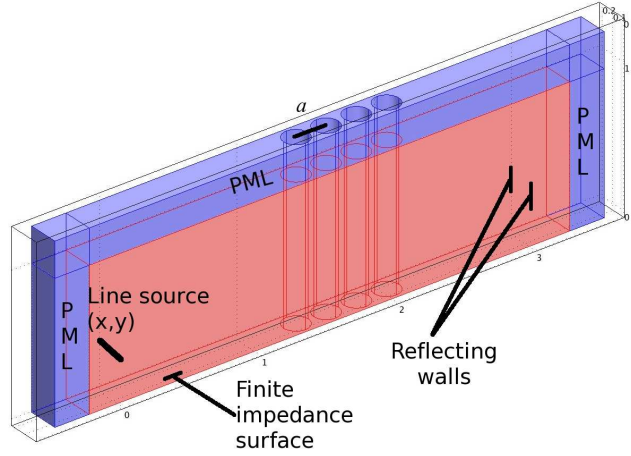}\\
(a)\\
\includegraphics[width=8.5cm]{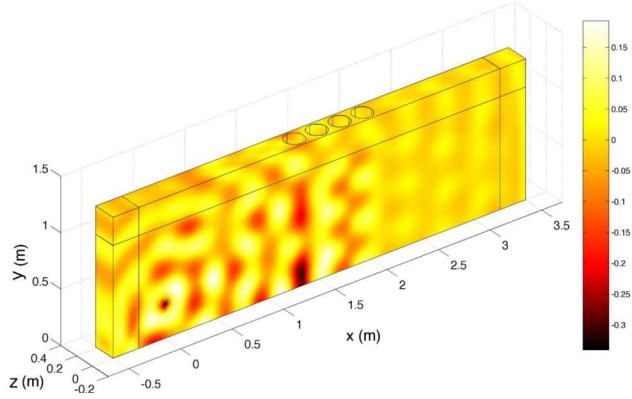}\\
(b) \end{center} \caption{\label{fig:figure} (Color online) (a) Numerical
solution domain. (b) Numerical simulation for the scattering problem
with completely rigid ground at 1000 Hz. The real value of the total pressure is plotted, $Re(P)$.}

\end{figure}

To numerically solve the problem we consider the geometry shown in fig. \ref{fig:figure}a. A line source is centred at point $\vec{r}_0=(x,y)$. The emitted acoustic field by this source at point $\vec{r}$ is $p_0=H_0(k|\vec{r}-\vec{r_0}|)$, where $H_0(x)$ is the Hankel
function of the first kind and zero order. A row of 4 scatterers separated by the lattice constant of the array, $a$, is enclosed between two completely reflecting walls which are parallel to the axis of the scatterers. These walls are also separated by a lattice constant, $a$. It is worth noting that the incident wave is not reflected by these walls, however the scattered waves from the cylinders are reflected by the walls reproducing the effect of a semi infinite periodic slab of 4 rows of scatterers. Then a reduced volume is used here to reproduce the effect of a semi infinite structure. This procedure is only valid for plane waves and line sources emitting cylindrical waves.

The commercial software COMSOL 3.5.a is used in this work to obtain the numerical predictions. The discrete domain where the solution is calculated is surrounded by Perfectly Matched Layers (PML) \cite{Berenguer94} which are an efficient alternative for emulating the Sommerfeld radiation condition in the numerical solution the scattering problems. The finite impedance surface where the cylinders are hold is modelled using the impedance boundary condition. We solve the problem for the scattered waves as the incident wave $p_s$ is itself solution of the wave equation. Thus, the total acoustic field is obtained adding the incident wave to the scattered one, then the PML only acts on the scattered wave. The solved problem has 10$^5$ degrees of freedom and 83.7 10$^4$ elements have been used to solve it. An example of the numerical solution at 1000 Hz is shown n fig. \ref{fig:figure}b.

The situation without the SC is very well known: the total field can be characterized by both the incident and the reflected waves. This last one can be obtained from the reflection coefficient of the finite impedance surface and the image method. Then, without the SC, the total acoustic field is \cite{Romero11}
\begin{eqnarray}
p_{tot}=p_0+p_s
\end{eqnarray}
where $p_0$ is the incident wave and $p_s=R(\vec{r}_0, \vec{r}; \nu)H_0(k|\vec{r'}-\vec{r'}_0|)$ represents the reflected one being $R(\vec{r}_0, \vec{r}; \nu)$ the reflection coefficient of the finite impedance surface. On the other hand, the parameter used in this work to characterize the attenuation properties of the complete system (SC and finite impedance surface) is the Insertion Loss (IL) defined here as the
difference between the sound level recorded with and without the SC
considering always the presence of the finite impedance surface. Then, 
\begin{eqnarray}
IL=20\log\frac{|p_{tot}|}{|P|}
\label{eq:IL}
\end{eqnarray}
where $P$ is the total acoustic field with both the surface and the SC.
We would like to remark that we have taken as reference for the IL the situation where the surface is present. 

Sometimes the dispersion relation of the periodic system is useful to compare the effect of the finite size on the scattering problems. Thus, the bands structures of the designed SC have been also
calculated using the PWE. 961 plane waves have been used in the calculation as this number produced a good convergence of the method for the case analysed here. We fix our attention in the range of frequencies included in the first pseudogap due to
the fact that this region has important relevance in the field of
the acoustic barriers which is perhaps the most exploited
application of SC for audible noise in airborne transmission.

\section{Results and discussion}
\label{sec:results}

\begin{figure}[hbt]
\begin{center}
\includegraphics[width=125mm]{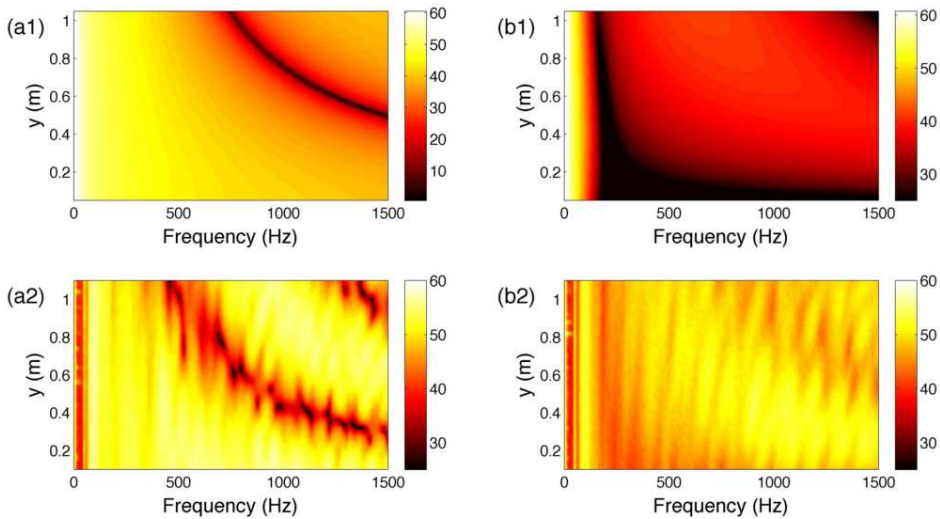}
\end{center} 
 \caption{\label{fig:figure3} (Color online) EA Theoretical analysis
in free field conditions for a rigid surface (a1) and finite
impedance surface (b1). EA Experimental measurements performed with
3DReAMS for the two previous cases: rigid surface (a2) and finite
impedance surface (b2). The geometrical conditions of both the
theoretical and the experimental results have been defined in Section
\ref{sec:tools}}
\end{figure}

First of all we analyze the case of the finite impedance surface without the SC. Fig. \ref{fig:figure3} shows both numerical predictions (1) and experimental results (2) of the dependence of the excess attenuation (EA) on the height of the measurement point and the frequency for the
rigid (a) and absorbent (b) surfaces. In the EA regions a destructive
interference between the direct and the reflected waves on the
surfaces is produced, and as a consequence an attenuation region is
observed. Note that the position of these regions depends on both the height and on the frequency.


On the other hand, in fig. \ref{fig:figure3} one can check the fairly good agreement
between the numerical and the experimental results, as a proof of
the accurate precision of our experimental set up. The existing
differences in the maps for the case of rigid surface (wood) can be
explained in terms of the assumption of acoustically rigid surface in the
numerical predictions. In the case of the finite impedance surface, EA
appears at lower values of both frequencies and heights than in the case of
rigid ground.

\begin{figure}[hbt]
\begin{center}
\includegraphics[width=105mm]{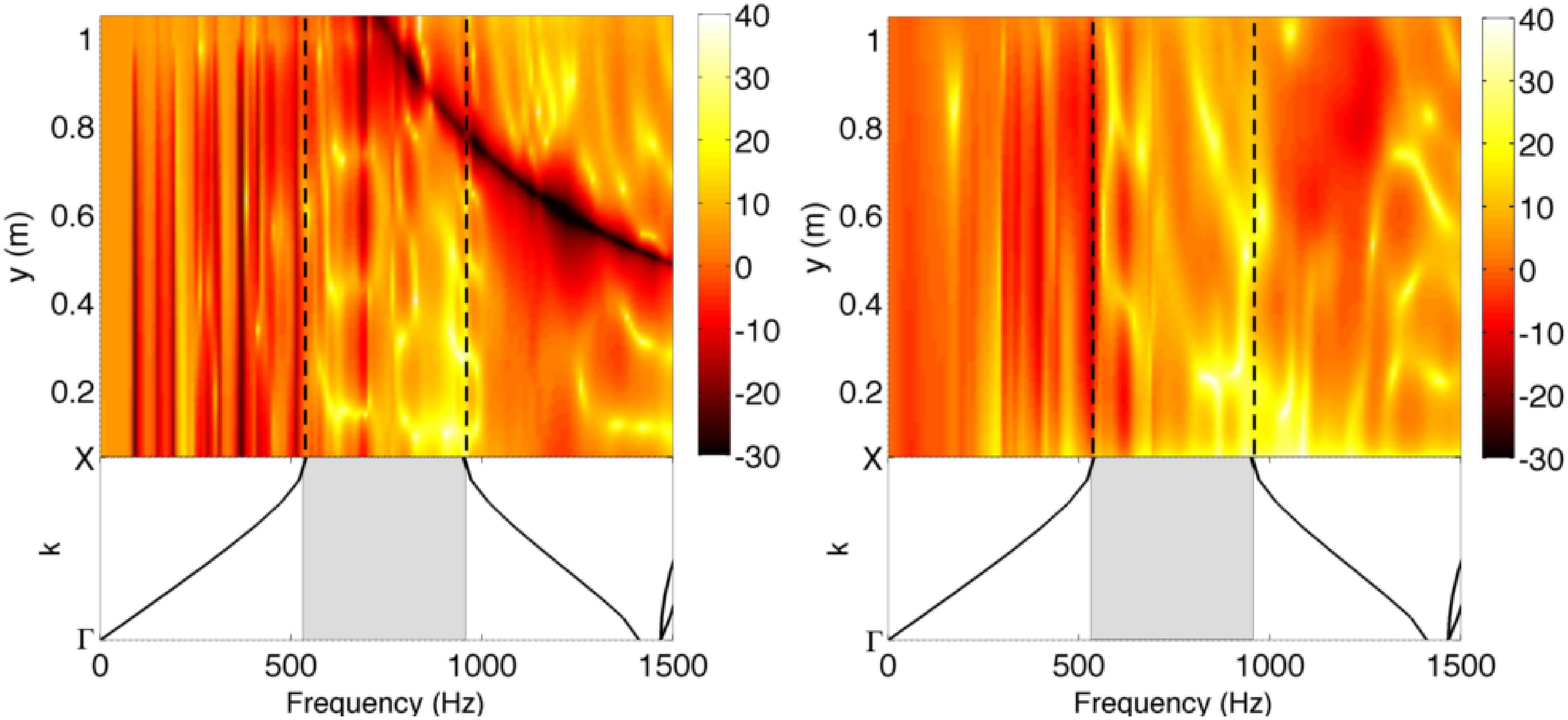}\\
(a1)\hspace{4cm}(b1)\\
\includegraphics[width=105mm]{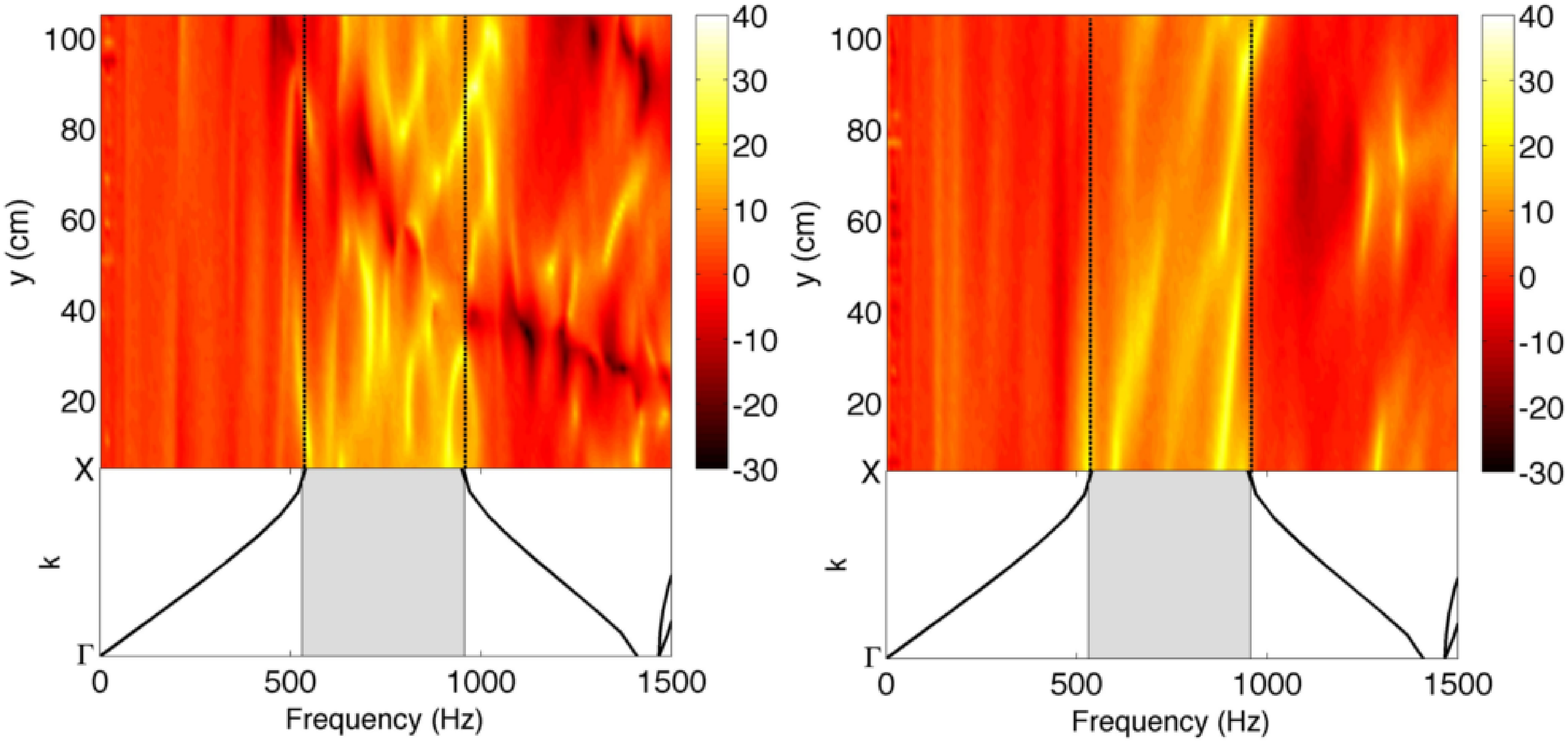}\\
(a2)\hspace{4cm}(b2)
\end{center} 
\caption{\label{fig:figure4} (Color online) IL theoretical analysis
of the SC-rigid surface interaction (a1) and SC-finite impedance
surface interaction (b1). Experimental results obtained with 3DReAMS
for the two previous cases: SC-rigid surface (a2) and SC-finite
impedance surface (b2). The results are obtained with the
geometrical conditions explained in Section II.B. The bands
structure has been calculated using the PWE method, with the next
parameters: $\rho_{cylinders}$=2700 kg/m$^3$, $\rho_{air}$=1.3
kg/m$^3$; $c_{cylinders}$=6400 m/s and $c_{air}$=340 m/s, and with 961 plane waves. We have focused our
attention in the properties of the first band gap at $\Gamma$X,
between 500 and 980 Hz}
\end{figure}

Once the EA effect of the surface is characterized we analyze the interaction between SC and both the rigid and the absorbent surfaces. Figs. \ref{fig:figure4}a1 and
\ref{fig:figure4}a2 show the numerical predictions and the experimental results respectively for the case of the acoustically rigid surface. In this
figures one can see that the maximum values of the IL appear
inside the range of frequencies corresponding to the first band gap
at $\Gamma$X direction, represented in the figures with the vertical
dotted lines as a prolongation of the edges of the bandgap shown in the band structures calculated using PWE method. For this case the interaction produces a positive
effect on the EA of the surface, that means, an increasing of the
attenuation properties of the surface. On the other hand, there are
some regions on the map where the IL is negative for both several
heights and frequencies, which mean a reinforcement of the
pressure level compared to the case of rigid surface alone which
is considered as the reference. Note that this reinforcement is
produced by the combined effect of the SC and the surface, and it means that the
attenuation level is lower than in the case of the rigid surface
alone, although some attenuation still exists compared with the free
field conditions.

Similar effects can be seen in figs. \ref{fig:figure4}b1 and
\ref{fig:figure4}b2 for the case of
finite impedance surface. The range of frequencies where the IL is
higher is included in the limits of the band gap at $\Gamma$X
direction, marked with vertical dashed lines.
Moreover, some regions in the map show a reinforcement of the
acoustical level, although here the size of these areas is lower
than in the rigid surface case. All these results are in good agreement
with the theoretical results showed in \cite{Romero11, Krynkin11} for the
SC-surfaces interaction in the two-dimensional case (horizontal cylinders).

\begin{figure}[hbt]
\begin{center}
\includegraphics[width=65mm]{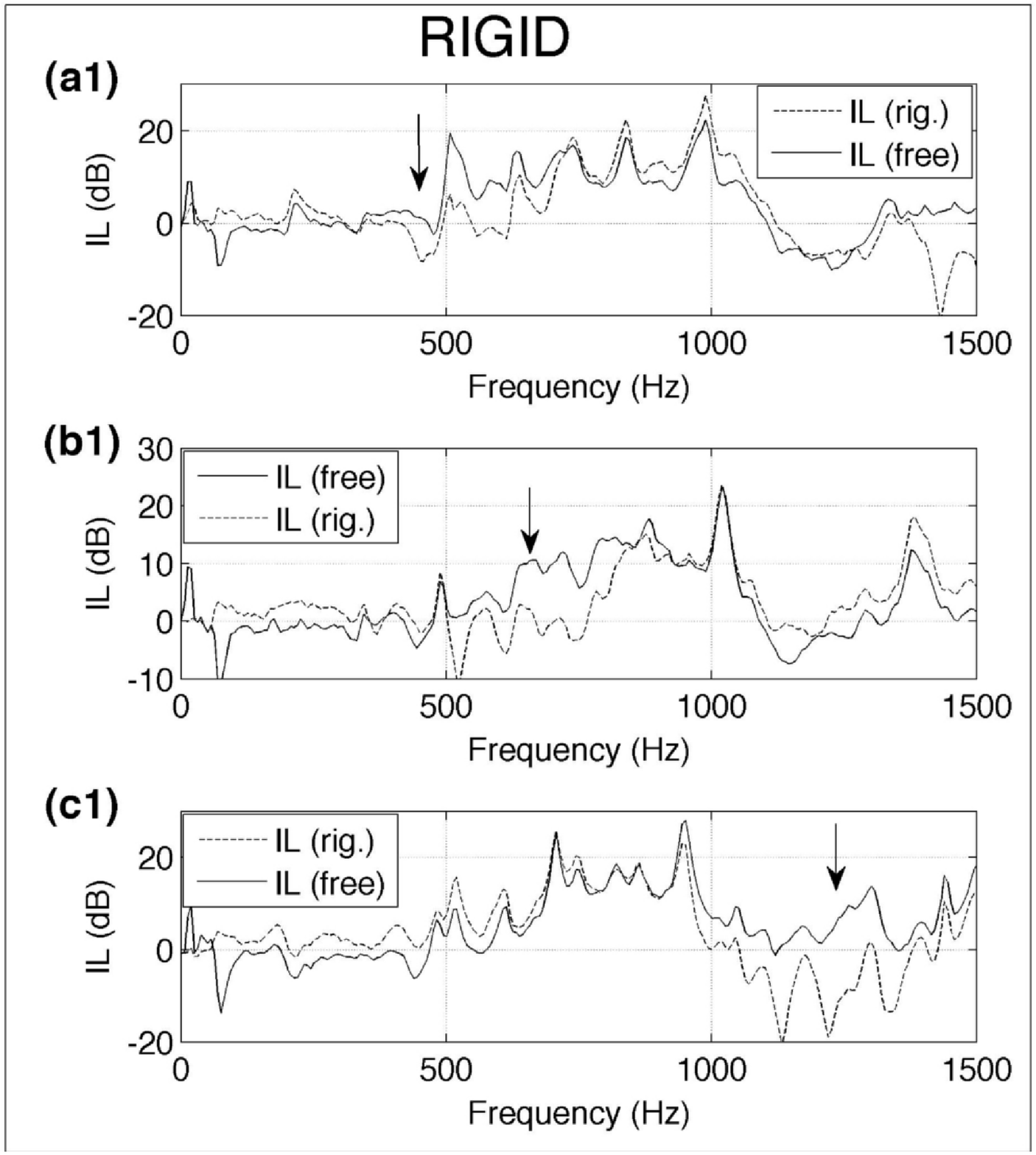}
\includegraphics[width=65mm]{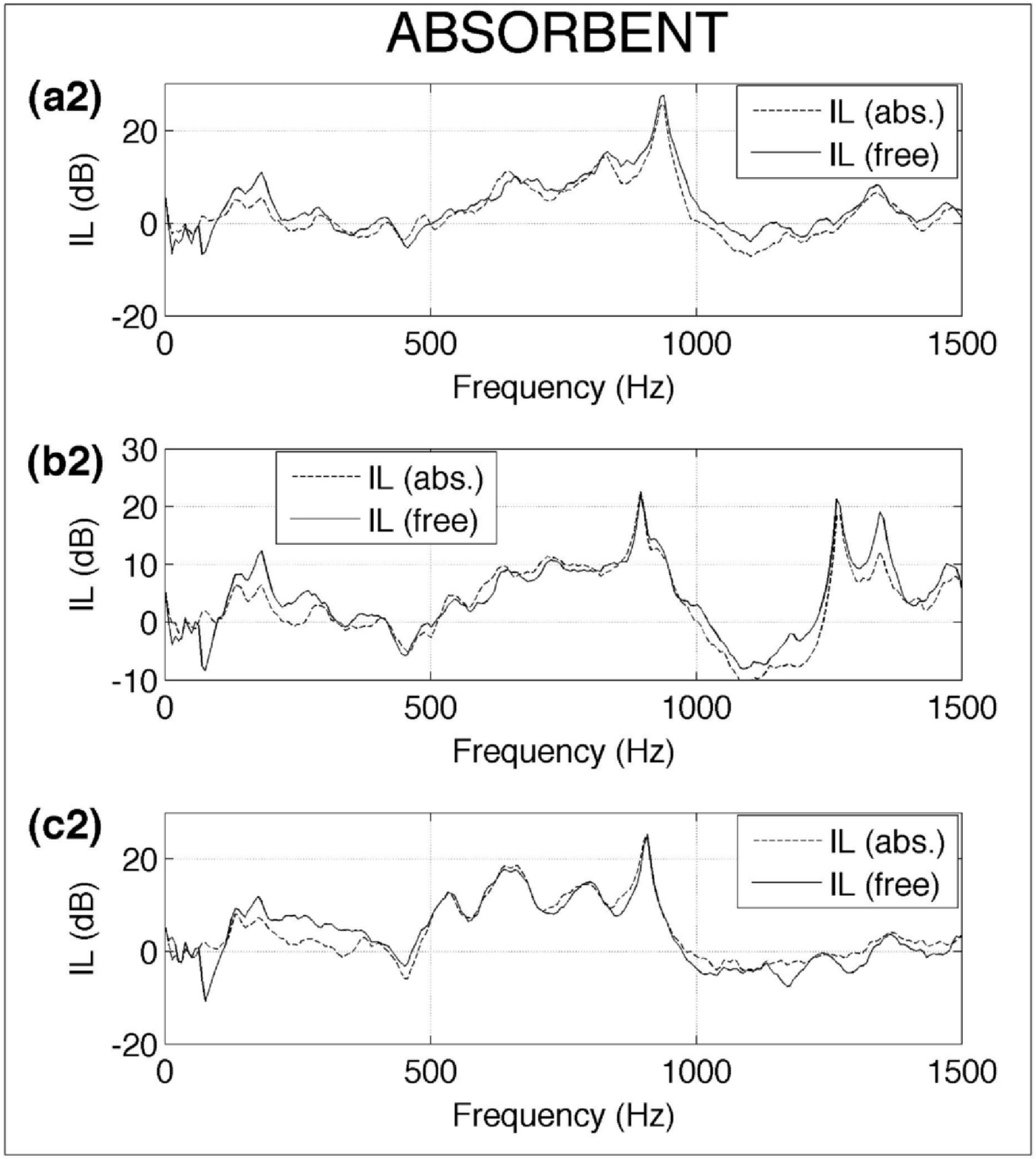}
\end{center} 
\caption{\label{fig:figure5} Experimental IL spectra
for three heights of the microphone (y=91 cm (a); y=65 cm (b) and  y=31 cm (c)). Continuous and dashed lines represent the IL of SC without groud and with ground respectively. Figure marked with 1 (2) represents the measurements for rigid (absorbent) ground. One can see in
the case of the rigid surface its effect on the attenuation
properties of SC (marked with arrows). In the case of a finite
impedance surface the existence of the surface does not affect the
SC attenuation properties.}
\end{figure}

Up to now, we have compared the dependence of the EA on the different
types of surfaces when a SC is introduced. In figure 5 we show the
opposite analysis that explains the influence of the surfaces in the
attenuation properties of SC. In this figure, one can compare the IL
spectra taking as reference the free field or the corresponding
surface for three heights of the microphone and for the two surfaces
analyzed. The dashed lines in fig. \ref{fig:figure5} show the IL calculated considering $p_{tot}$ in eq. \ref{eq:IL} the acoustic field considering only the source and the finite impedance surface. The continuous lines represent the IL calculated considering $p_{tot}$ the corresponding for the free field (without the surface).  In the case of rigid surface (figs. \ref{fig:figure5}a1, \ref{fig:figure5}b1 and \ref{fig:figure5}c1) one
can see the increasing of the attenuation properties of the SC when
the rigid surface is considered. In the case of the finite impedance
surface there is not a relevant variation in the size of the
$\Gamma$X pseudogap when the surface is considered.

\section{Concluding remarks}
\label{sec:conclusions}
We have shown both numerically and experimentally the interaction between the attenuation properties of both a SC and a finite impedance surface. We have focused our analysis on the case where the axis of the rigid
cylinders that form the SC are perpendicular to the surfaces. For
this realistic case we have found a good agreement between both the
numerical predictions and the experimental data measured using
3DReAMS. Both the EA, due to the presence of the finite impedance
surface, and the band gap of the SC interact in order to show a combined
effect that depends on the frequency and on the relative position
between the measuring point and the position of the source. We have
analyzed this interaction under the point of view of the attenuation properties of the SC, finding that for the case of rigid
surfaces the combined effect is relevant and it has to be taken into
account. However, for finite impedance surfaces with high absorption values, the interaction
practically disappears and the attenuation properties of the SC in
the measuring points are only dependent on the periodicity of the
array as well as on the number of scatterers. The results shown in
this work can be useful to design attenuation devices based on
periodic materials as SC acoustic barriers as well as for the design
of experiments in which the effect of the surface could interact
with the transmission properties of SC.

\begin{acknowledgments}
This work was supported by MEC (Spanish Government) and FEDER funds,
under grands MAT2009-09438.
\end{acknowledgments}

\end{document}